\documentclass[preprint,12pt]{elsarticle}

\usepackage{amssymb}
\usepackage[symbol]{footmisc}
\usepackage{orcidlink}

\usepackage{graphicx}%
\usepackage{multirow}%
\usepackage{amsmath,amssymb,amsfonts}%
\usepackage{amsthm}%
\usepackage{mathrsfs}%
\usepackage[title]{appendix}%
\usepackage{xcolor}%
\usepackage{textcomp}%
\usepackage{manyfoot}%
\usepackage{booktabs}%
\usepackage{algorithm}%
\usepackage{algorithmicx}%
\usepackage{algpseudocode}%
\usepackage{listings}%
\usepackage[autostyle]{csquotes}

\journal{...}

\begin{document}

\begin{frontmatter}



\title{Viscoelastic amplification of the pull-off stress in the detachment of a rigid flat punch from an adhesive soft viscoelastic layer}


\author[inst1]{{Ali} {Maghami} }

\affiliation[inst1]{organization={{Department of Mechanics, Mathematics and Management}, {Politecnico di Bari}},
            addressline={via Orabona 4}, 
            city={Bari},
            postcode={70125}, 
            country={Italy}}

\author[inst1,inst2]{{Michele} {Ciavarella} }

\author[inst1,inst2]{{Antonio} {Papangelo}  \footnote[1]{Corresponding author\\ Email address: antonio.papngelo@poliba.it ({Antonio} {Papangelo}).}}

\affiliation[inst2]{organization={{Department of Mechanical Engineering}, {Hamburg
University of Technology}},
            addressline={{Am Schwarzenberg-Campus 1}}, 
            city={Hamburg},
            postcode={21073}, 
            country={Germany}}

\begin{abstract}
{The problem of the detachment of a sufficiently large flat indenter from a plane adhesive viscoelastic strip of thickness ``{\textbf{\textit{b}}}'' is studied. For any given retraction speed, three different detachment regimes are found: (i) for very small ``{\textbf{\textit{b}}}'' {the detachment stress is constant and equal to the theoretical strength of the interface,} 
(ii) for intermediate values of ``{\textbf{\textit{b}}}'' the detachment stress decays approximately as {\textbf{\textit{b}}}$^{-1/2}$, (iii) for thick layers a constant detachment stress is obtained corresponding to case the punch is detaching from a halfplane. By using the boundary element method a comprehensive numerical study is performed which assumes a linear viscoelastic material with a single relaxation time and a Lennard-Jones force-separation law. Pull-off stress is found to consistently and monotonically increase with unloading rate, but to be almost insensitive to the history of the contact. Due to viscoelasticity, unloading at high enough retraction velocity may allow punches of macroscopic size to reach the theoretical strength of the interface. Finally, a corrective term in Greenwood or Persson theories considering finite size effects is proposed. Theoretical and numerical results are found in very good agreement.}
\end{abstract}




\begin{keyword}
Viscoelasticity \sep Crack propagation \sep Thin layer \sep Detachment force.
\end{keyword}

\end{frontmatter}


\section{Introduction}\label{sec1}

Soft materials are of great interest in the scientific community as for
their applicability in many engineering fields ranging from the automotive
sector \cite{Lorenz2015}
, biomechanics \cite{Mazzolai2019}, soft robotics \cite{Gio2021a,Gio2021b},
manipulators \cite{Shintake}, tires grip
\cite{Lorenz2015}, sealing of syringes
\cite{Huon2022}, finger-touch-pad interactions \cite%
{Forsbach2023,Felicetti2023}
, soft tissue adhesion for regenerative medicine \cite%
{Burke2007,Gumbiner1996}
and pressure-sensitive adhesives \cite{Asbeck}. With soft polymers
macroscopic adhesion due to van der Waals adhesive interactions remains
strong \cite{Dahlquist}, whereas in hard materials it is easily canceled by
the inevitable surface roughness.

In many of these applications the bond strength is a crucial mechanical
property and it is often quantified by measuring the apparent adhesion
strength as given by the maximum pulling force per unit area in a tensile
bond test \cite{peng2020decohesion}. Peng et al. \cite{peng2020decohesion} have already elucidated how the critical pull-off force of a flat rigid axisymmetric punch adhered to an \textit{elastic} film of finite thickness depends by two dimensionless parameters.
The former shows a transition from uniform detachment (DMT-like) to
crack-like propagation (JKR-like), while the latter is a correction factor
due to finite thickness of the film. 

However, in many of the applications mentioned above, soft
materials (polymers, elastomers) are employed, which are known to be viscoelastic, i.e. they
exhibit a frequency-dependent modulus and dissipation \cite{christensen2012theory}, and this complicates their mechanical adhesive behaviour. Numerous experiments in steady state conditions have shown that the apparent surface energy $\Delta \gamma $, i.e. the energy per unit area needed to separate two ideally parallel surfaces, during the crack opening
is related to the crack speed $V$ through a power law function \cite{Creton,
Barquins1981, Gent, gentpetrich}, commonly referred as the Gent and Schulz
empirical law \cite{GentSchultz} 
\begin{equation}
\frac{\Delta \gamma \left( V\right) }{\Delta \gamma _{0}}=1+\left( \frac{V}{%
V_{ref}}\right) ^{n},  \label{GS}
\end{equation}%
where $\Delta \gamma _{0}$ is the adiabatic surface energy (or thermodynamic
work of adhesion), $V$ is the crack velocity, $V_{ref}=\left(
ka_{T}^{n}\right) ^{-1}$ and $k,n$ are constants with $0<n<1$ and $a_{T}$ is
the WLF factor to translate viscoelastic modulus results at various
temperatures $T$ \cite{Williams}. In its simplest form Gent and Schulz
empirical law Eq. (\ref{GS}) is generally a good phenomenological model for
opening cracks, while for closing cracks a reduced apparent surface energy $%
\Delta \gamma $ is observed which generally shows the reciprocal of that law 
\cite%
{greenwood2004theory,Persson2005,persson2017,Persson2021,Schapery,SchaperyII}%
. There is no indication in the law of a limit enhancement, nor it is clear
how far it can be used for transient conditions. Finally, and most
importantly, in this empirical law, there is no indication on possible size
effects, i.e. on how the parameters of the law should be affected by
geometry, and $\frac{\Delta \gamma \left( V\right) }{\Delta \gamma _{0}}>1$
which instead we shall find is not always true, even for advancing crack.

There are two main approaches which have been attempted to capture more {%
fundamentally the propagation of viscoelastic crack propagation}. One is
based on the description of the processes occurring at the crack tip through
a cohesive zone model \cite{Greenwood1981, Schapery, SchaperyII,
Schapery2022}, and as such is rather general as it can take into account of
initiation of the crack, transient propagation, and steady state. Also it
may show a transition to a cohesive rupture for small enough cracks (what
Peng et al. \cite{peng2020decohesion} call uniform DMT-like detachment in their case, see also \cite{Papangelo2023,AffVio2022, VioAffRange}), in principle it may be
generalized to rate-dependent cohesive laws, and to non linear materials.
In practice, it attempts to model the real processes occurring at the crack
tip. The other, developed by Persson and coauthors \cite{Persson2005,
persson2017, Persson2021},  takes an \textquotedblleft
energy-based\textquotedblright\ approach and is restricted to linear materials and to steady
state conditions. It is derived by equating the power input in the system with the power dissipated by viscoelastic losses and by the rate at which energy is spent to create 
new surfaces. The energy-based approach finds different results for finite
size systems where it seems to show non-monotonic $\Delta \gamma \left(
V\right) /\Delta \gamma _{0}$ and also $\Delta \gamma \left( V\right)
/\Delta \gamma _{0}>1$ \cite{persson2017}, which is in contrast to the limit
case of cohesive failure where $\Delta \gamma \left( V\right) /\Delta \gamma
_{0}$ can decrease down to zero for very small crack (see \cite{Papangelo2023}), as we shall discuss in details here with respect to our geometry. {Only }for a semi-infinite system and linear material both approaches {yield a very similar monotonically increasing behavior} of $\Delta \gamma \left( V\right) /\Delta \gamma _{0}$ with {respect to} $V$ up
to the theoretical \textquotedblleft high-frequency\textquotedblright\ limit
of $\Delta \gamma /\Delta \gamma _{0}=E_{\infty }/E_{0}$, {where} $E_{\infty
}$ and $E_{0}$ respectively {represent} the glassy (high frequency) and the
rubbery (low frequency) modulus of the viscoelastic material \cite%
{ciavacricrimcmeek}\footnote{%
In their basic form, both approaches consider $\Delta \gamma _{0}$ to be an
intrinsic material property and therefore rate-independent, as we shall also
assume here. }. \ Furthermore, the cohesive model has been applied modelling
of bi-materials crack (one elastic, one viscoelastic) \cite{CPM2021} showing transient dissipation can be arbitrarily large while loads remain finite and hence dissipation should not be taken as an indication of true fracture energy \cite%
{ciavarella2022transient}.

Recently, it has been shown that, depending on the indenter geometry, the
loading history may or may not affect the detachment force. For a Hertzian
indenter, Afferrante and Violano \cite{AffVio2022, VioAffRange} have shown
that the loading velocity and the maximum indentation reached during the
loading phase can significantly influence the pull-off force. One the other
hand, Papangelo and Ciavarella \cite{Papangelo2023} {found that when
considering} an axisymmetric flat punch, the loading history has a
relatively weak effect, and the primary determinant of the pull-off force is
largely the unloading velocity. This is partly related to the fact that,
for a given preload, the contact area {achieved} at maximum indentation
strongly depends on the loading history only for a Hertzian indenter, while
it remains fixed for a flat punch. {This interpretation is somehow confirmed
by the recent work of Muller et al. \cite{muller2023revealing} which further
considers the contact of a flat punch indenter with superimposed small scale
roughness. Their study focuses on the significant hysteresis that is
observed during crack closure and opening which is obtained by the
concurrent presence of viscoelasticity and adhesion. They find that small
scale roughness can indeed leads to loacal jumps -in and -out of contact,
which causes the dependence of the detachment force on the preload.}

While previous works have focused on the detachment from viscoelastic
semi-infinite substrates \cite%
{Papangelo2023,afferrante2023exploring,forsbach2023simple}, the influence of
the layer thickness on the detachment process has been mostly overlooked.
Nevertheless, the latter has a broad interest in engineering applications
where often a thin layer of viscoelastic material is used, as in Figure \ref%
{fig:application}, as well as to exploit the thin layer testing geometries,
like in Peng et al. \cite{peng2020decohesion} which could permit to extract cohesive
properties as well as surface energy properties. 
\begin{figure}[t]
\centering
\includegraphics[width=4.5in]{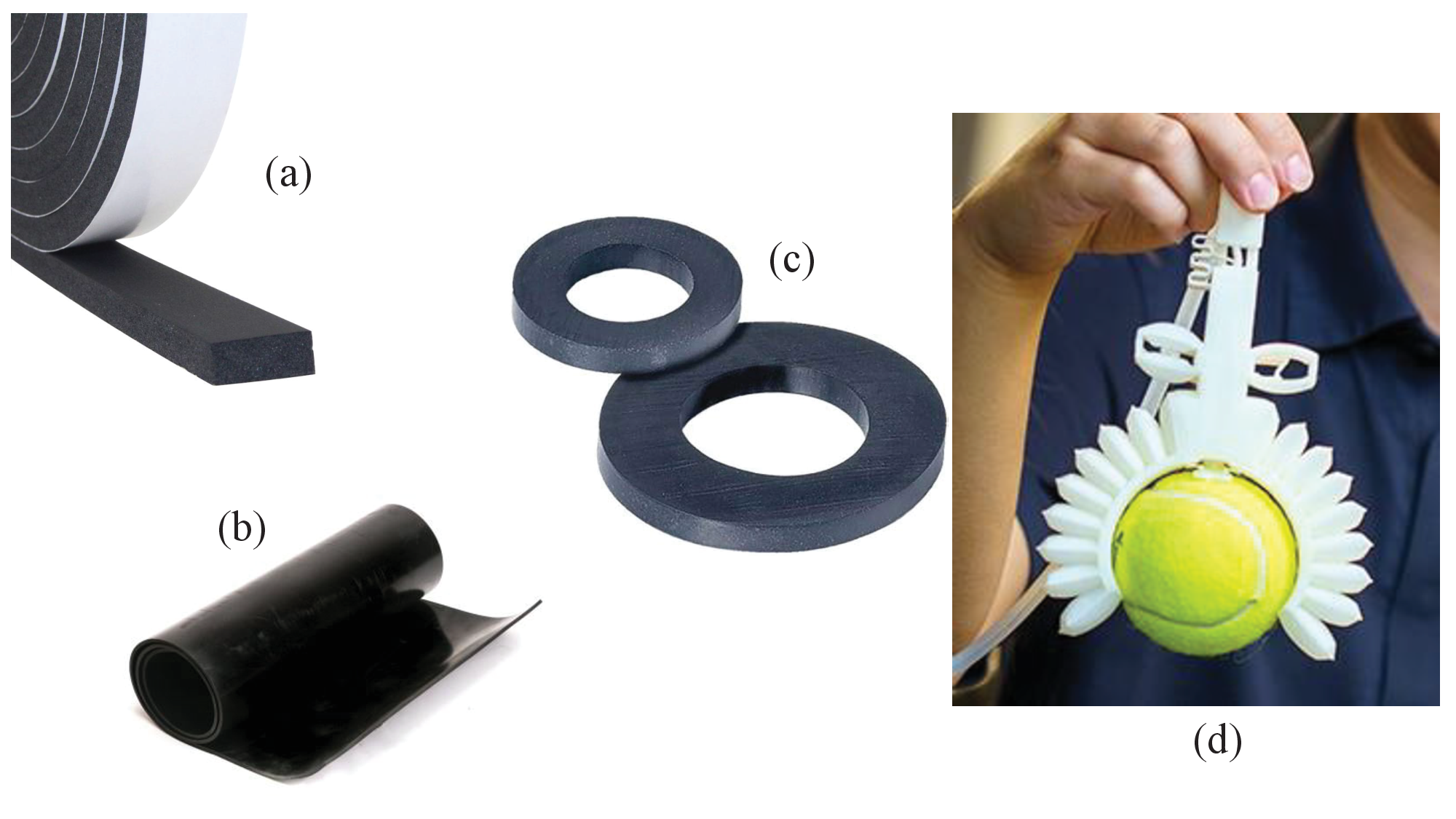}
\caption{{Practical engineering applications where thin polymeric layers are
used: (a) rubber tape, (b) polymeric carpet, (c) gasket, (d) soft gripper
for delicate handling and manipulation (UCSD Jacobs School of Eng., CC
BY-NC-SA 4.0).}}
\label{fig:application}
\end{figure}

Hence, in the present study, we focus on the detachment problem of a flat
punch indenting a thin adhesive viscoelastic layer of finite thickness $b$
in plane strain conditions. The
remainder of the paper is organized as follows. In Section 2 the elastic
solution for halfplane contact is recalled. In Section 3, the case of a thin
layer is considered leveraging on the the \textquotedblleft thin
strip\textquotedblright\ assumption by Johnson \cite{Johnson1985} and
applying the Griffith energy balance. This serves to determine the limiting
solutions within which the viscoelastic results should be confined. In
Section 4 the numerical scheme is introduced which is based on the boundary
element method and it assumes a standard linear model for the viscoelastic
layer and a Lennard-Jones force-separation law for the contact interactions,
which is rate-independent. In Section 5 the numerical results are presented
and compared with Greenwood theory for viscoelastic crack propagation \cite%
{greenwood2004theory} which we extend for finite size effects. In Section 6
the conclusions are drawn. The study is carried on with particular emphasis
on the effect of (i) the loading history, (ii) the layer thickness and (iii)
the unloading velocity on the pull-off force and on the effective adhesive
energy.  

\section{Detachment from a halfplane} \label{sec:2}
Let us consider the plane contact problem of a flat punch of semi-width $a$ indenting an elastic adhesive frictionless halfplane with Young modulus $E$ and Poisson ratio $\nu$. By applying the Griffith energy balance, the pull-off force \cite{Barber2018, Maugis2000} is given by
\begin{equation}
P_{po}=L\sqrt{2\pi E^{\ast}\Delta\gamma_{0}a}\;,%
\end{equation}
where $L$ is the layer width, $E^{\ast}=E/\left(  1-\nu^{2}\right)  $ is the plane strain elastic modulus. Hence the mean interfacial stress at pull-off is 
\begin{equation}
\overline{\sigma}_{po}=\sqrt{\frac{\pi E^{\ast}\Delta\gamma_{0}}{2a}}\;,\label{shp}%
\end{equation}
which has the classical Linear Elastic Fracture Mechanics (LEFM) square-root dependence with respect to the punch semiwidth $a$. Overbar indicates here
the mean value. This implies {that} smaller punches have a higher pull-off stress, potentially reaching the theoretical strength (or the cohesive strength) of the interface, denoted as $\sigma_{0}$. {This is typically observed for punches with a semi-width less than the following typical fracture length}
\begin{equation}
a_{0}=\frac{\pi}{2}\frac{E^{\ast}\Delta\gamma_{0}}{\sigma_{0}^{2}}\;.%
\end{equation}

In what follows in the paper {we shall assume that the punch size is }
$a>>a_{0}$ as we are interested in the transitions due to the finite size of the layer
rather than the size of the punch. For the latter effect the reader is referred to Ref. \cite{Papangelo2023}.

{Hence}, in dimensionless form, {we have the following relations}%

\begin{equation}
\widehat{\overline{\sigma}}_{po}=\frac{\overline{\sigma}_{po}}{\sigma_{0}}=\frac{1}{\sqrt{a/a_{0}}}\;,
\end{equation}
\begin{equation}\label{eq:a0}
\frac{a_{0}}{h_{0}}=\frac{9\sqrt{3}\pi}{32\Sigma_{0}}\approx\frac
{1.53}{\Sigma_{0}}\;,
\end{equation}
where we have assumed a Lennard-Jones force-separation law, for which $\Delta\gamma_{0}=\alpha_{LJ} h_{0}\sigma_{0}$, {where} $h_{0}$ {is} the range of interaction and $\alpha_{LJ} =\frac{9\sqrt{3}}{16}$ is a constant,  $\widehat{a}=a/h_{0}$, $\Sigma_{0}=\sigma_{0}/E^{\ast}$ is usually
in the range {of} {$\left[  0.1\div1\right]$} for soft polymers \cite{Jagota2000, Tang2006, Maugis2000}, implying $a_{0}$ to be 1 to 10 times higher than the range of attractive
forces. For a true crystal, this would imply a range of few nanometers, but
for soft materials, the range of adhesive forces may be larger. If a viscoelastic material with relaxed modulus $E_{0}$ and instantaneous modulus $E_{\infty}$ is considered, then in the limit of very slow and very fast unloading rate we will have
\begin{equation}
\left\{
\begin{array}
[c]{cc}%
\widehat{\overline{\sigma}}_{po}=\sqrt{\frac{9\sqrt{3}\pi}{32\Sigma_{0}\widehat{a}}%
}=\frac{1}{\sqrt{a/a_{0}}}\;; & \qquad\text{slow limit }E=E_{0}\;,\\
\widehat{\overline{\sigma}}_{po}=\sqrt{\frac{9\sqrt{3}\pi}{32\Sigma_{0}k\widehat{a}}%
}=\frac{1}{\sqrt{ka/a_{0}}}\;; & \qquad\text{fast limit }E=E_{\infty}\;,%
\end{array}
\right.
\end{equation}
where $k=E_{0}/E_{\infty}$ and $\widehat{a}=a/h_0$.

\section{Detachment from a thin layer} \label{sec:3}
{If the substrate has a finite thickness, it is {necessary} to consider the effect of thickness in the analysis. Hence, here we focus on the plane contact problem of a flat punch with a semi-width of $a$ indenting an adhesive layer with a thickness of $b$ (Fig. \ref{fig:identer}). We first focus on the linear elastic solution, and then we will provide the limiting solutions for the viscoelastic problem based on the elastic formulation.}

\subsection{Elastic layer}
Let us consider the layer in plane strain and supported by a rigid foundation. In the following the case of frictionless contact between the layer and the rigid substrate is considered while the correction due to the Poisson effect for the case of a layer perfectly bonded to the substrate is shown in the Appendix-I.
\begin{figure}[h]
\centering
\includegraphics[width=4.5in]{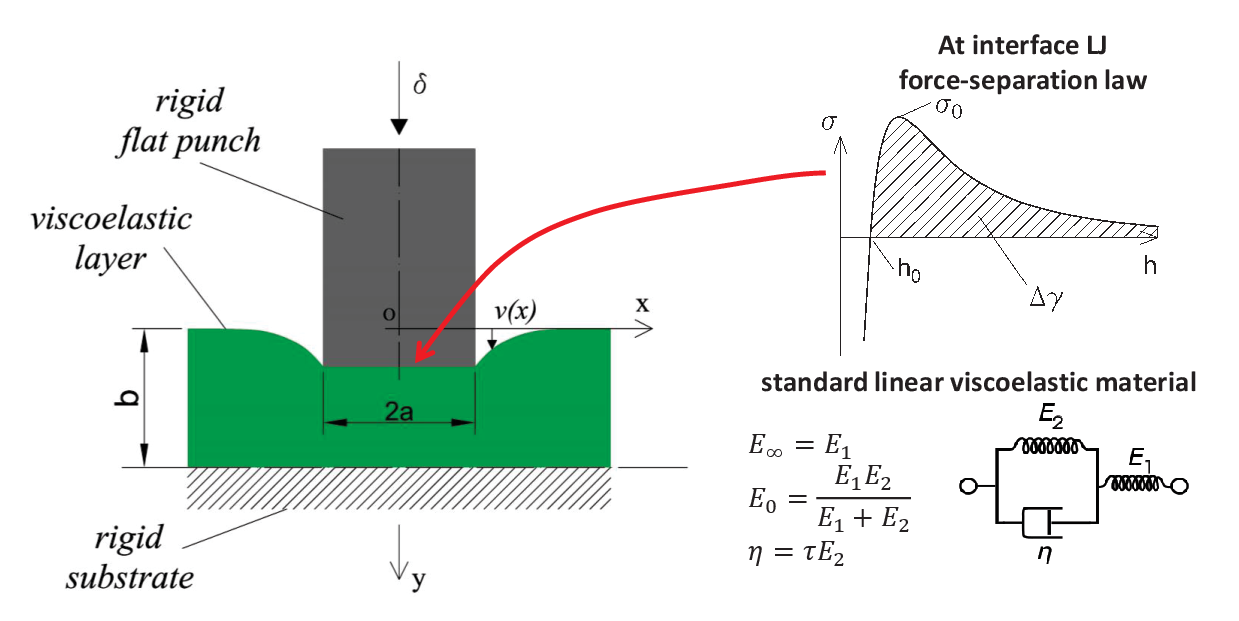}
\caption{On the left is a sketch of a flat punch being loaded on a viscoelastic adhesive layer of thickness $b$. The numerical implementation employs the Lennard-Jones force-separation law, while the viscoelastic material is represented using a standard linear model, as depicted in the lower-right panel.}
\label{fig:identer}
\end{figure}

Following Johnson \cite{Johnson1985}, we assume {that} plane sections remain plane upon loading. Hence for the case of no friction between the layer and the rigid substrate, the load $P$ and the corresponding elastic strain energy $U_{E}$ stored in the layer are
\begin{align}
P &  =-2aLE^{\ast}\frac{\delta}{b}\;,\\
U_{E} &  =aLE^{\ast}\frac{\delta^{2}}{b}\;,%
\end{align}
where $\delta$ is the indentation considered positive when {the flat punch is} approaching the substrate, consequently, $P$ {is} positive when tensile. At unloading, the Griffith energy balance requires the elastic strain energy released per unit area to be equal to the surface energy. Hence, assuming
detachment occurs immediately, {we have the following relations

\begin{equation}
\frac{1}{2L}\frac{\partial U_{E}}{\partial a}=\Delta\gamma\rightarrow\left\{
\begin{array}
[c]{c}%
\delta_{po}=-\sqrt{\frac{2b\Delta\gamma}{E^{\ast}}}\;,\\
\overline{\sigma}_{po}=\sqrt{\frac{2E^{\ast}\Delta\gamma}{b}}\;,%
\end{array}
\right.  \label{sl}%
\end{equation}
where, $\delta_{po}$ and $\overline{\sigma}_{po}$ are the indentation and the
average interfacial stress at pull-off respectively. Notice that the pull-off stress depends on the layer thickness as $\overline{\sigma}_{po}\propto b^{-1/2}$, hence it is possible to define a characteristic thickness $b_{0}$ of the substrate where
$\overline{\sigma}_{po}$ reaches the theoretical interfacial strength $\sigma_{0}$, i.e.%

\begin{equation}\label{eq:b0}
b_{0}=\frac{2E^{\ast}\Delta\gamma}{\sigma_{0}^{2}}=\frac{4}{\pi}a_{0}%
\approx1.27a_{0}\;,%
\end{equation}
which is of the same order of magnitude of $a_{0}$. Experiments with PDMS
elastomers in Peng et al. \cite{peng2020decohesion}, show that this $b_{0}$ is of the order of $0.1$ mm, where clearly their loading rate corresponds to a certain effective
elastic modulus. On the other {hand, in the limit of a very thick layer, we should obtain the half-plane solution, for which we can utilize Eq. (\ref{shp}) and Eq. (\ref{sl}) to determine a length scale}

\begin{equation}\label{eq:b1}
b_{1}=\frac{4a}{\pi} \;,%
\end{equation}
{with the meaning that for substrates thicker than $b_1$ one should anticipate the half-plane behaviour.} Notice that, while $b_{0}$ is a characteristic lengthscale that depends on the material and interfacial properties, $b_{1}$ depends on the
punch semi-width. Overall {as indicated by Eq. (\ref{eq:b0}) and Eq. (\ref{eq:b1}), and as illustrated in Fig \ref{fig:regimes}, }we {identify} the following three regimes%

\begin{equation}
\left\{
\begin{array}
[c]{cr}%
\overline{\sigma}_{po}=\sigma_{0}\;, & \qquad b<b_{0}\\
\overline{\sigma}_{po}=\sqrt{\frac{2E^{\ast}\Delta\gamma}{b}}\;, & \qquad b_{0}\leq b\leq
b_{1}\\
\overline{\sigma}_{po}=\sqrt{\frac{\pi E^{\ast}\Delta\gamma}{2a}}\;, & \qquad b>b_{1}%
\end{array}
\right.
\end{equation}
or, in dimensionless form,
\begin{equation} \label{eq:regimes}
\left\{
\begin{array}
[c]{cr}%
\widehat{\overline{\sigma}}_{po}=1\;, & \qquad b/a_{0}<\frac{4}{\pi}\\
\widehat{\overline{\sigma}}_{po}=\sqrt{\frac{9\sqrt{3}}{8\Sigma_{0}\widehat{b}}%
}=\sqrt{\frac{4}{\left(  b/a_{0}\right)  \pi}}\;, & \qquad\frac{4}{\pi}\leq
b/a_{0}\leq\frac{4}{\pi}\frac{a}{a_{0}}\\
\widehat{\overline{\sigma}}_{po}=\sqrt{\frac{9\sqrt{3}\pi}{32\Sigma_{0}\widehat{a}}%
}=\frac{1}{\sqrt{a/a_{0}}}\;. & \qquad b/a_{0}>\frac{4}{\pi}\frac{a}{a_{0}}%
\end{array}
\right.
\end{equation}

\subsection{Limiting solutions for a viscoelastic layer}
Let us assume that the layer is constituted by a viscoelastic material with relaxed Young modulus $E_{0}$ and instantaneous Young modulus $E_{\infty}$ so that $k=E_{0}/E_{\infty}$. In the limit of very slow/very fast unloading rate the substrate behaves as elastic. Thus, for the case of no friction between the substrate and the layer, {according to Eq. (\ref{sl}), one can anticipate the following two scenarios}%
\begin{equation}
\left\{
\begin{array}{cc}
\overline{\sigma}_{po}=\sqrt{\frac{2E_{0}^{\ast}\Delta\gamma}{b}}\;, & \qquad\text{\enquote{very slow}}\\
\overline{\sigma}_{po}=\sqrt{\frac{2E_{\infty}^{\ast}\Delta\gamma}{b}}\;, & \qquad\text{\enquote{very fast}}
\end{array}
\right.
\end{equation}
or in dimensionless form 
\begin{equation} \label{eq:fastandslow}
\left\{
\begin{array}{cc}
\widehat{\overline{\sigma}}_{po}=\sqrt{\frac{4}{(b/a_{0})\pi}}\;, & \qquad\text{\enquote{very slow}}\\
\widehat{\overline{\sigma}}_{po}=\sqrt{\frac{4}{k(b/a_{0})\pi}}\;, & \qquad\text{\enquote{very fast}}
\end{array}
\right.
\end{equation}
{where, one should notice that} for rapid unloading {(very fast scenario)} the pull-off stress will reach the cohesive strength by the following value of the substrate thickness:
\begin{equation}
{b_{0\infty}}=\frac{4{a_{0}}}{k\pi} \label{b0inf}\;.%
\end{equation}
{However, } for a thick layer, the halfplane limit will be always obtained {at $b_{1}=4a/\pi,$ irrespective of the unloading rate}. Figure \ref{fig:regimes} schematically displays the elastic limits at low and high retraction speed that constitute the bounds for the possible viscoelastic solutions.

Form Eq.s (\ref{eq:b1}) and (\ref{b0inf}) it follows that if $a/a_{0}<E_{\infty}/E_{0}$ then $b_{0\infty}>b_{1}$. {In other words,} if $a<a_0/k$ at a high enough retraction velocity, {it is possible to reach} the adhesive strength of the interface. It is easy to find elastomers with {$E_{\infty}/E_{0}\simeq10^{3}\div10^{4}$} \cite{Bonfanti2020}. This
implies that punches with semiwidth $a$ much larger than $a_{0}$ can still
reach the theoretical interfacial strength if unloading is performed fast
enough. Also, this holds for layer thickness. If we assume that the
experiments with PDMS in Peng et al. \cite{peng2020decohesion} were conducted
relatively slowly, when $b_{0}$ was of the order of $0.1$ mm, then clearly ${%
b_{0\infty }}$ \ could reach very large values, provided the punch is also
large enough. This peculiarity may be exploited in the future as a technique
to improve/enhance interfacial adhesion in micro-structured interfaces by
optimizing not only the micro-pillar geometry, but also the unloading
protocols. 
\begin{figure}[h]
\centering
\includegraphics[width=4.5in]{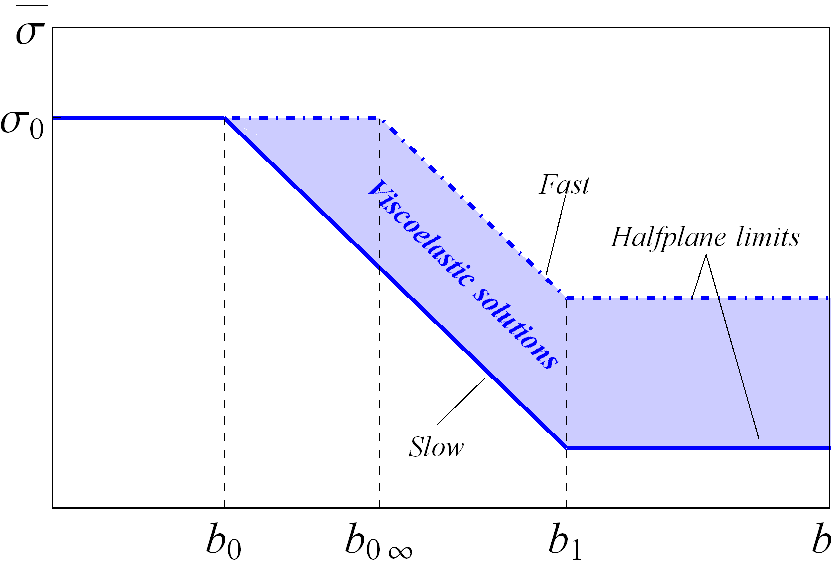}
\caption{Schematic representation of the three possible detachment
regimes. }
\label{fig:regimes}
\end{figure}

\section{Numerical implementation of the adhesive viscoelastic contact
problem}  \label{sec:4}
In this section, the numerical scheme used to solve the adhesive viscoelastic contact problem is introduced. We utilized the boundary element method, which necessitates the discretization solely of the interface. A similar code has been used in previous works for solving both adhesive elastic \cite{PapCia2020} and viscoelastic \cite{Papangelo2023} axisymmetric contact problems, therefore, in this section, we will focus on the essential adaptations required to tailor the implementation for a plane viscoelastic strip.

The flat punch interacts with the viscoelastic layer {according to the} Lennard-Jones 3-9 force-separation law {defined as}%
\begin{equation}
\sigma\left(  h\right)  =\frac{8\Delta\gamma}{3h_{0}}\left[  \left(
\frac{h_{0}}{h}\right)  ^{3}-\left(  \frac{h_{0}}{h}\right)  ^{9}\right]\;,
\label{LJ}%
\end{equation}
where $\sigma$ is the traction ($\sigma>0,$ when it is tensile), $h$ is the interfacial gap and $h_{0}$ is the equilibrium distance. The theoretical strength of the interface ({maximum tensile stress}) is then {equal to} $\sigma_{0}=\Delta\gamma/(\alpha_{LJ}h_0)$
and takes place at a separation {equal to} $h=3^{1/6}h_{0}$. The gap {is a} function {of the in-plane coordinate $x$} as%
\begin{equation}
h(x)=-\delta+h_{0}+v\left(  x\right)\;,  \label{h}%
\end{equation}
where, $v\left(  x\right)$ is the deflection of the viscoelastic layer with respect to the origin $\left(0,0\right)  $ ($v\left(x\right)$ is positive as shown in   Fig. \ref{fig:identer}). Here, Eq. (\ref{h}) is solved numerically in a discrete manner at the $N=M+1$ nodes, being $M$ the number of equally spaced elements {with the length of $c=2a/M$.}  
Following Bentall and Johnson \cite{Bentall1968} we implemented the method of overlapping triangles, i.e. for the $n$-th node the pressure is 0 at node $x_{n-1}$, rises linearly to $p_{n}$ at node $x_{n}$ and then falls linearly to 0 at node $x_{n+1}$, which gives overall a linear variation of the contact pressure $p\left(  x\right)$ over the considered domain. With respect to the case of constant pressure elements, a picewise-linear distribution of normal tractions produces a displacement field which is everywhere smooth and continuous. Hence, according to Bentall and Johnson \cite{Bentall1968} the vertical deflection at node $m$ of an elastic layer relatively to the origin $\left(  x,y\right)  =\left(  0,0\right)  $ due to a triangular distribution of pressure centered in $x_{n}$ is%
\begin{equation}
v_{m}=aB\frac{4}{\pi E^{\ast}}\left[  I_{A0}+I_{A}\left[  m-n\right]
+4zI_{AR}\left[  m-n\right]  \right]  p_{n}\label{vm2}\;,%
\end{equation}
where $\left\{  m,n\right\}  $ are integers numbers, $p_{n}$ is the pressure acting on the $n$-th node determined using Eq. \ref{LJ} ( $p_{n}>0$ when it is tensile), $B=b/a$, $z=c/4b=1/2BM$ and $I_{A0}$, $I_{A}$, $I_{AR}$ are the
following integral functions\footnote{Care should be taken when integrating
$I_{A0}$ which converges slowly. The Appendix 3 of Bentall and Johnson \cite{Bentall1968} suggests a convenient integration strategy we have also adopted. Notice that Bentall and Johnson 
 \cite{Bentall1968} contains a misprint as the second part in which
$I_{A0}$ is split up should be integrated over the interval $\left[
\delta,+\infty\right]  $.}%
\begin{align}
I_{A0} &  =\frac{2}{z}\int_{0}^{\infty}\left(  \frac{1-\cosh\beta}{\beta
+\sinh\beta}\right)  \frac{\sin^{2}\left(  \beta z\right)  }{\beta^{3}}%
d\beta\;,\\
I_{A}\left[  m-n\right]   &  =-\frac{4}{z}\int_{0}^{\infty}\left(
1+\frac{1-\cosh\beta}{\beta+\sinh\beta}\right)  \frac{\sin^{2}\left(  \beta
z\right)  }{\beta^{3}}\sin^{2}\left(  \beta z\left(  m-n\right)  \right)
d\beta\;,\\
I_{AR}\left[  m-n\right]   &  =\int_{0}^{\infty}\frac{\sin^{2}\left(
\eta\right)  \sin^{2}\left(  \eta\left(  m-n\right)  \right)  }{\eta^{3}}%
d\eta,\qquad\eta=\beta z\;.
\end{align}
By applying the superposition principle, the normal deflection $v_{m}$ at node $m$ due to a piecewise linear distribution of pressure can be written as
\begin{equation}
v_{m}=
\frac{1}{E^{\ast}}{\displaystyle\sum\limits_{n=1}^{N}}
G_{mn}p_{n},%
\end{equation}
where {each} column of the compliance matrix $\{(1/E^{*})\overline{\overline{G}}\}_{NxN}$
corresponds to the displacement field due to a unity triangular pressure centered at node $n$ being all the other nodes unloaded. Therefore, the displacement field and, correspondingly, the compliance matrix can be readily computed using Eq. (\ref{vm2}).  {Once} the elastic solution {is obtained}, the displacement field of the viscoelastic layer ${v}\left(  x,t\right)  $, can be determined by the
elastic-viscoelastic correspondence principle in the form of Boltzmann
integrals \cite{Chri} as 
\begin{equation}
{v}\left(  x,t\right)  {=\frac{1}{E_{0}^{\ast}}\int G}{\left(
x,s\right)  \int_{-\infty}^{t}c(t-\tau)\frac{dp(s,\tau)}{d\tau}d\tau ds}\;,
\end{equation}
where $c(t)$ is the dimensionless creep compliance function, the strain variation after an application of a constant unit stress, and, in our discrete formulation, the Green function ${G}{\left(x,s\right) }$ is replaced by the appropriate tensor $\{\overline{\overline{G}}\}_{NxN}$, so that the viscoelastic nodal displacements $\{{v}(t)\}_{Nx1}$ at time $t$ are%
\begin{equation}\label{vt}
\{{v}(t)\}_{Nx1}  \boldsymbol{=}\{\overline{\overline{G}
}\}_{NxN}\ast \biggl\{ \frac{1}{E_{0}^{\ast}}{\int_{-\infty}^{t}c(t-\tau)\frac{d p}{d\tau}d\tau} \biggl\}_{Nx1}, 
\end{equation}
where the symbol \enquote{$\ast$} stands for the row by column product. For the linear viscoelastic material, the standard model is assumed with a single relaxation time $\tau$, composed by a spring placed in series with an element constituted by a dashpot and a spring in parallel (see Fig. \ref{fig:identer}), for which the dimensionless creep compliance function is as follows%
\begin{equation}
c(t)=\left[  1+\left(  k-1\right)  \exp\left(  -\frac{t}{\tau}\right)
\right]\;.
\end{equation}

being $\tau$ the relaxation time of the material. Hence, by using a sequential time-marching continuation, we solved Eq.s (\ref{LJ},\ref{h},\ref{vt}), where at each time step an iterative scheme is used to determine the equilibrium solution. 

\section{Results} \label{sec:results}
Here the results of the numerical investigations are shown by using the following dimensionless parameters
\begin{equation}
\widehat{a}= \frac{a}{h_0}; \;\; \;\; \widehat{\sigma}(x)= \frac{\sigma(x)}{\sigma_0};\;\;\;\; \widehat{\overline{\sigma}}= \frac{P}{2aL \sigma_0}; \;\;\;\;\widehat{\delta}= \frac{\delta}{h_0}; \;\;\;\;\widehat{t}= \frac{t}{\tau},
\end{equation}
{and $\widehat{\overline{\sigma}}_{po}$} is the ({dimensionless}) average stress at pull-off and is defined as $\widehat{\overline{\sigma}}_{po}= \textrm{max}(\widehat{\overline{\sigma}})$. If not stated differently, in our simulations we considered $M=200$, $\Sigma_0=0.05$, and $k=0.1$.

\subsection{History dependence}

{Viscoelastic materials typically exhibit a ``{{history-dependent}}'' response and this tremendously affects the detachment force in Hertzian indenters \cite{AffVio2022, VioAffRange}. Hence, first we aimed to explore how different loading scenarios affect the detachment characteristics of the flat indenter we considered, while keeping the unloading rate constant. The simulations were carried out under displacement control using a trapezoidal function (see inset in Fig. \ref{fig:loadeffect}.a) which main parameters are shown in Table \ref{tab:my_table}. We define the dwell time as $ \widehat{t}_{dwell} =\widehat{t}_2-\widehat{t}_1$, the unloading rate $\widehat{r}=(\widehat{\delta}_{load}-\widehat{\delta}_{unload})/(\widehat{t}_3-\widehat{t}_2)$, and the loading rate as $\widehat{r}_{load}=(\widehat{\delta}_{load}-\widehat{\delta}_{0})/(\widehat{t}_1)$ with reference to the  the inset of Figure \ref{fig:loadeffect}.a. We endeavored to investigate the impact of different loading protocols meticulously although the unloading curves presented in Figure \ref{fig:loadeffect} are restricted to the six different protocols described in Table \ref{tab:my_table}.}
\begin{table}[]
    \centering
        \caption{Description of parameters governing the loading protocols for the curves presented in Figs. \ref{fig:loadeffect} and \ref{fig:loadeffect2.} .}
    \label{tab:my_table}
\begin{tabular}{c c c c c c }
\hline
 {Loading protocol} & $\widehat{\delta}_0$ & $\widehat{\delta}_{load}$ & $\widehat{r}_{load}$ & $\widehat{r}$ & $\widehat{t}_{dwell}$ \\
\hline
{\#} 1 & 1 & 1 & very {fast} & 10 & 0 \\
{\#} 2 & 1 & 1 & very {slow} & 10 & 0 \\
{\#} 3 & 0 & 1 & 5 & 10 & 0 \\
{\#} 4 & 2 & 2 & very {fast} & 10 & 0 \\
{\#} 5 & 2 & 2 & very {slow} & 10 & 0 \\
{\#} 6 & 0 & 2 & 5 & 10 & 0 \\
\hline
\end{tabular}
\end{table}

\begin{figure} 
     \centering
     \includegraphics[width=4.5in]{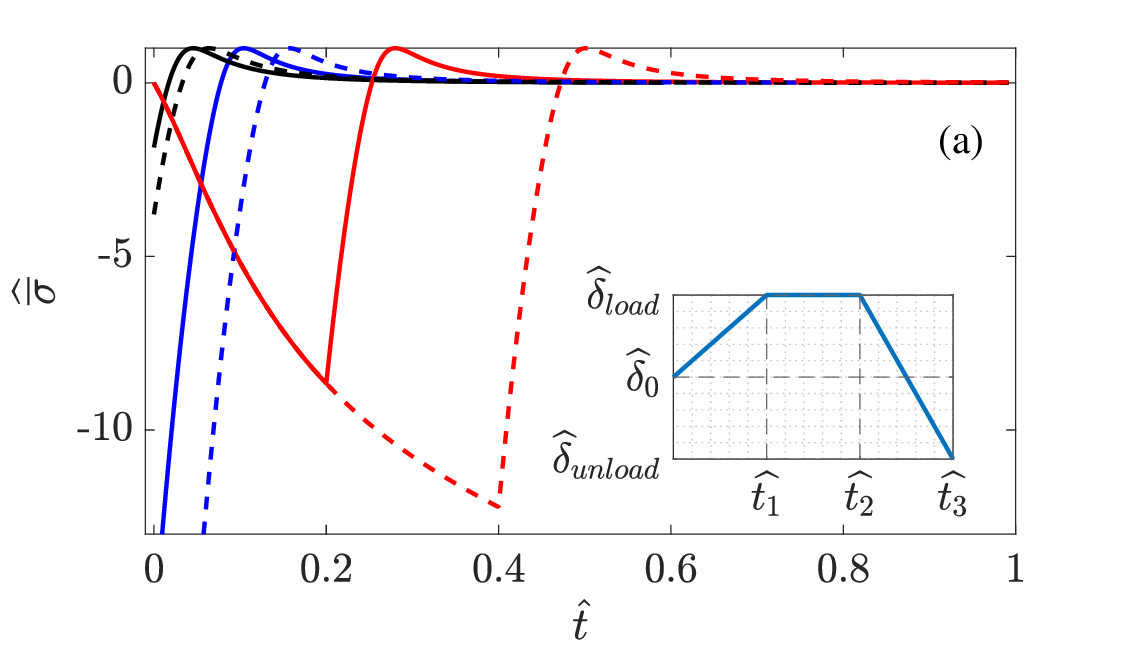}
     \includegraphics[width=4.5in]{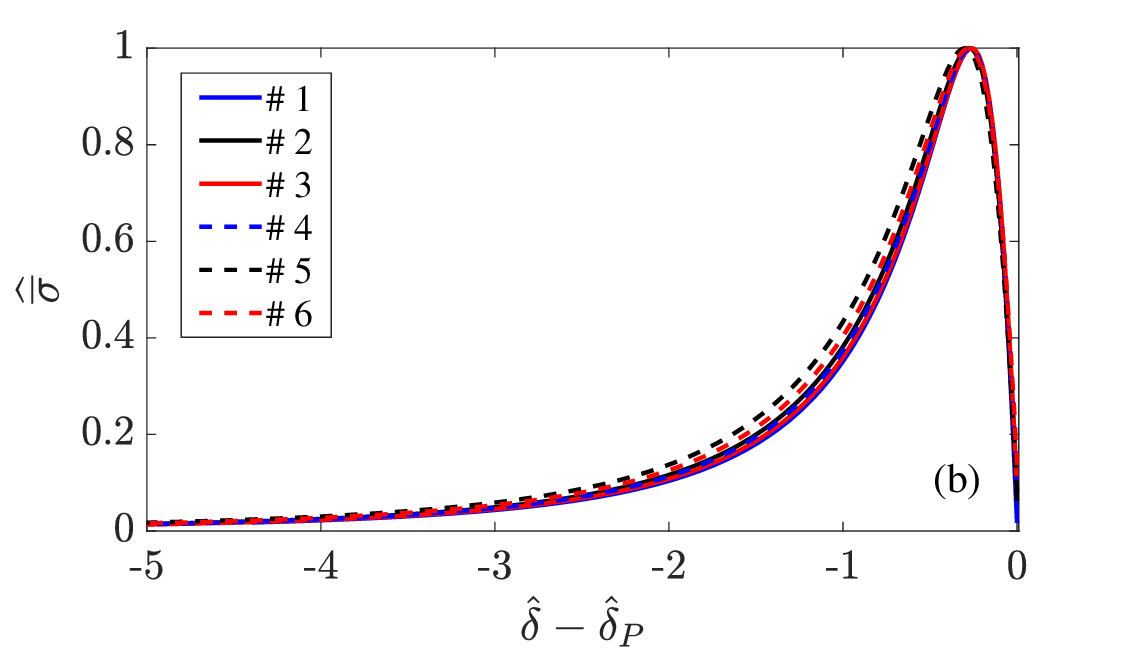}
              \caption{(a) Unloading curves for $\Sigma_0 = 0.05, k = 0.1$ and punch of radius  $\widehat{a}/\widehat{a}_0=64.85$ from a fully relaxed viscoelastic surface with different loading protocols. (b) The identical curves displayed in (a) are reiterated here subsequent to a horizontal axis shift equal to $\widehat{\delta}_{P=0}$.} \label{fig:loadeffect}
\end{figure}

{These loading scenarios included: (i) Unloading from a fully relaxed substrate after a slow loading process, indicated by the black curves (2,5). (ii) Unloading following rapid loading, causing the substrate to exhibit an elastic response with \(E(t = 0) = E_{\infty}\), as denoted by the blue curves (1,4).
(iii) Unloading after indenting the substrate at a constant loading rate \(\widehat{r}_{load} = 5\), represented by the red curves (3,6). It's important to note that while the loading phase is not shown for curves (1, 2, 4, 5), we accounted for the pre-loading effect in our simulations. Furthermore, the maximum indentation depth \(\widehat{\delta}_{load}\) was kept at \(\widehat{\delta}_{load} = 1\) for curves (1, 2, 3) (solid lines) and set to \(\widehat{\delta}_{load} = 2\) for curves [4, 5, 6] (dashed lines). The punch has $a/a_0=64.85$, and the viscoelastic layer's (dimensionless) thickness is $B=b/a=0.005$ which is equivalent to $\widehat{b}/\widehat{a_0}=0.3242$.}

\begin{figure}[h]
\centering
\includegraphics[width=4.5in]{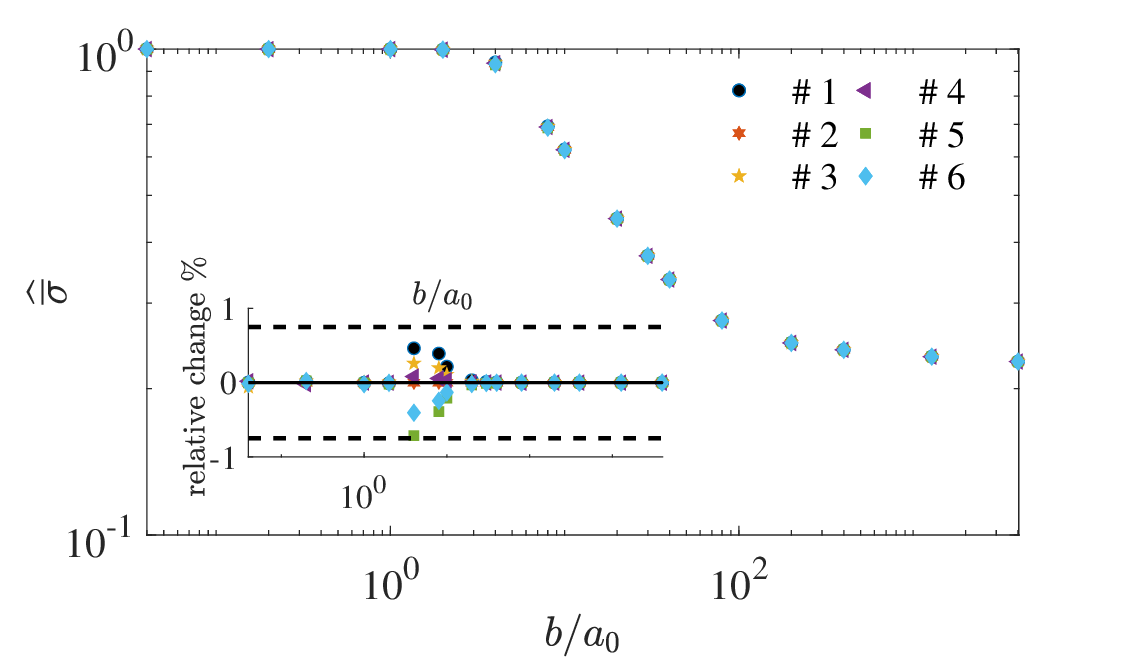}
\caption{Normalized pull-off stress as a function versus the normalized layer's thickness for different loading protocols.}
\label{fig:loadeffect2.}
\end{figure}

{The primary result from Fig. \ref{fig:loadeffect} is that, regardless of the significant variations in loading conditions, the magnitude of the pull-off stress remains consistent across various loading histories. Indeed, one can conclude that the pull-off stress remains nearly unaffected by the loading history. In Figure \ref{fig:loadeffect}.b, we present the same curves as in Fig. \ref{fig:loadeffect}.a, but with a shift in the horizontal axis by $\delta_{P}$. This shift corresponds to the indentation depth at which the normal load vanishes during unloading. It helps to better observe the slight changes in the unloading trajectories. Furthermore a comprehensive investigation on the effect of the different loading scenario was conducted for various layer thicknesses. In Figure \ref{fig:loadeffect2.}, we illustrate $\widehat{\overline{\sigma}}_{po}$ versus the variation in thickness $b/a_0$. The legend in Figure \ref{fig:loadeffect2.} clarifies that the plot presents results derived from all the loading conditions in Table \ref{tab:my_table}, all with the same unloading rate. Remarkably, these plots closely overlap, indicating very similar values across the various simulations for all the thicknesses tested. To further quantify the distinctions between these loading cases, we examined the relative changes in pull-off stress with respect to one specific case that serves as the foundation for our subsequent investigations. The results are plotted in the inset of Figure \ref{fig:loadeffect2.} as a relative change for various thickness values. It's evident that within a certain accuracy we can state that the detachment force of a flat indenter from a viscoelastic adhesive strip is \textit{negligibly influenced by the loading history of the contact}. For the remainder of the paper we will consider unloading the viscoelastic strip from a fully relaxed condition, unless explicitly stated otherwise, we assume $\widehat{\delta}_{0} = \widehat{\delta}^{{load}} = 1$, ${\widehat{t}_{dwell}} = 0$.}

\subsection{Dependence on the unloading rate}

{After establishing that the loading history does not influence the pull-off stress $\widehat{\overline{\sigma}}_{po}$, we examine how $\widehat{\overline{\sigma}}_{po}$ varies with {respect to} the layer thickness for four different unloading rates: $\widehat{r} = [0.1, 1, 10, 100]$ represented in Fig. \ref{fig:pulloffvsthickness} by black diamonds, green circles, red squares, and pink triangles, respectively. Figure \ref{fig:pulloffvsthickness} shows a comprehensive analyses for the punch semi-width $\widehat{a}/\widehat{a}_0=64.85$. The results are obtained starting from a fully relaxed substrate. For $b < b_0\simeq1.27a_0$, we reach the cohesive limit, where the pull-off stress remains independent on both the unloading rate and the layer thickness, approaching the theoretical value $\widehat{\overline{\sigma}}_{po} = 1$. Most importantly, for $b_0 < b < b_1$, the curves align well with the LEFM (Linear Elastic Fracture Mechanics) solution we have derived in Section \ref{sec:3} showing a scaling of $\propto b^{-1/2}$. Here, the pull-off stress increases with the unloading rate, and the pull-off data consistently stay well by the ``{slow}'' and ``{fast}'' limits we derived, represented by the blue dashed and solid black lines, respectively. For a thickness larger than $b_1$ the curves align with the half-plane solution and, for a given unloading rate, the pull-off stress remains constant independently on the layer thickness. 
\begin{figure}[h]
         \centering
         \includegraphics[width=4.5in]{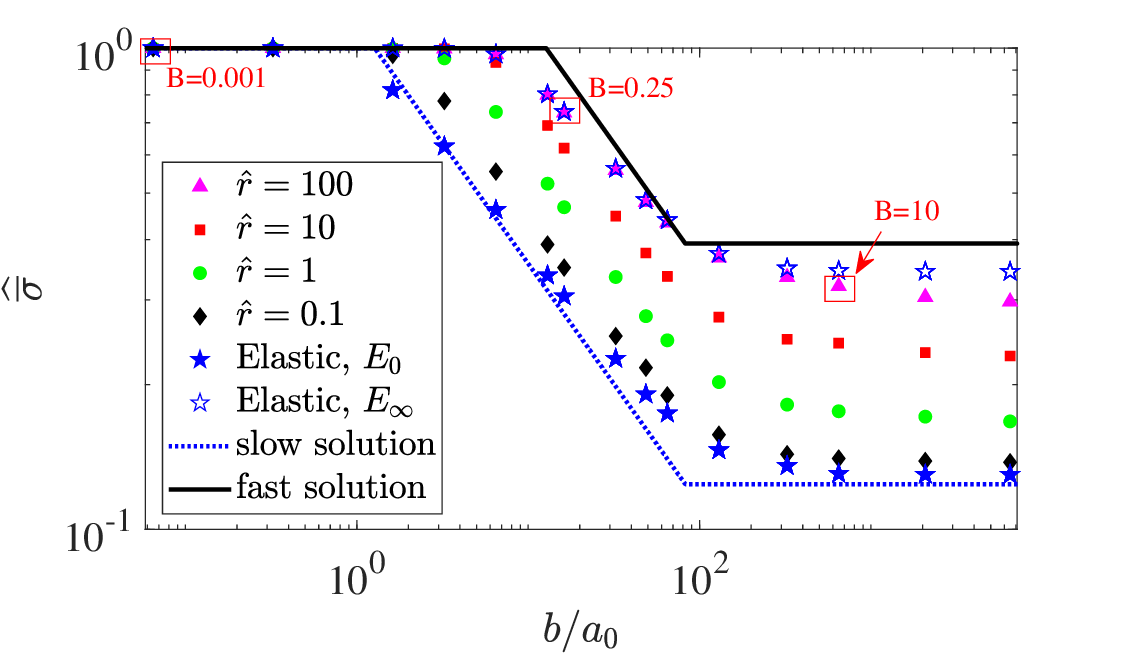}
         \caption{{Normalized pull-off stress as a function of the normalized layer's thickness for different unloading rates {of a punch with $\widehat{a}/\widehat{a}_0=64.85$.}} }\label{fig:pulloffvsthickness}
\end{figure}

Notice that in the theoretical elastic solution the detachment happens with no propagation ($a_c=a$, being $a_c$ the semi-width of the crack ligament). Clearly, this condition is never achieved in a more refined cohesive zone model, as it is the one we have implemented numerically. This accounts for the small deviations we found in the limiting case of very fast and very slow unloading between numerical and theoretical results. Nevertheless, to ascertain the correctness of the numerical viscoelastic results, the plots also include the curves obtained unloading an \textit{elastic} strip with modulus $E_0$ (filled blue stars) and $E_{\infty}$ (empty blue stars). One easily recognizes that the viscoelastic solutions are perfectly bounded between the two limiting elastic cases.}

{To support our conclusion we focus on the mechanism of crack propagation and stress distribution at the interface from the unloading onset up to pull-off.}
Figures \ref{fig:ptilde} show the stress distribution for three specific cases out of the 120 cases shown in Fig. \ref{fig:pulloffvsthickness}. All the cases are for unloading rate $\widehat{r}=100$. The punch radius in Figure \ref{fig:ptilde} is $\widehat{a}/\widehat{a}_0= 64.85$. The corresponding points for these three cases are highlighted with red squares in Fig. \ref{fig:pulloffvsthickness}. According to Fig. \ref{fig:ptilde}, during unloading, the crack propagates at the interface hence the semi-width of the crack ligament $a_c$ is smaller than the punch semi-width $a$ when pull-off happens. This explains the difference between the expected pull-off stress from the analytical limits and the actual pull-off stress. 

{Figure \ref{fig:ptilde} displays three distinct cases with different values of $B$ (0.001, 0.25, and 10), denoted as Figures \ref{fig:ptilde}a, \ref{fig:ptilde}b, and \ref{fig:ptilde}c, respectively. Figure \ref{fig:ptilde}a pertains to the cohesive zone, where the detachment occurs at $a_c\simeq a$ and with a uniform distribution of tensile tractions at the interface. For a more comprehensive understanding of crack propagation and the detachment process, we have included gap plots on the right side of Figure \ref{fig:ptilde}. These plots represent the gap $H(x)={h(x)}/{h_0}-1$ between the rigid punch and the viscoelastic layer as a function of the in plane coordinate $x$. Figure \ref{fig:ptilde}a illustrates that in the cohesive region the gap is uniformly distributed at the interface, with no crack formation.
This is, in fact, the reason why we can achieve $\widehat{\overline{\sigma}}_{po}$ numerically with the same results as the expected analytical results (see Figure \ref{fig:pulloffvsthickness} for $B=0.001$). In contrast, for the other two cases, as depicted in Figures \ref{fig:ptilde}b and \ref{fig:ptilde}c, we observe crack propagation, with detachment occurring at $a_c<a$. Notably, for larger values of $B$, a reduction in the ratio $a_c/a$ is evident, resulting in the small deviation observed in Fig. \ref{fig:pulloffvsthickness} between the numerical and the analytical results.}
\begin{figure}[H]
\begin{center} 
\includegraphics[width=4.5in]{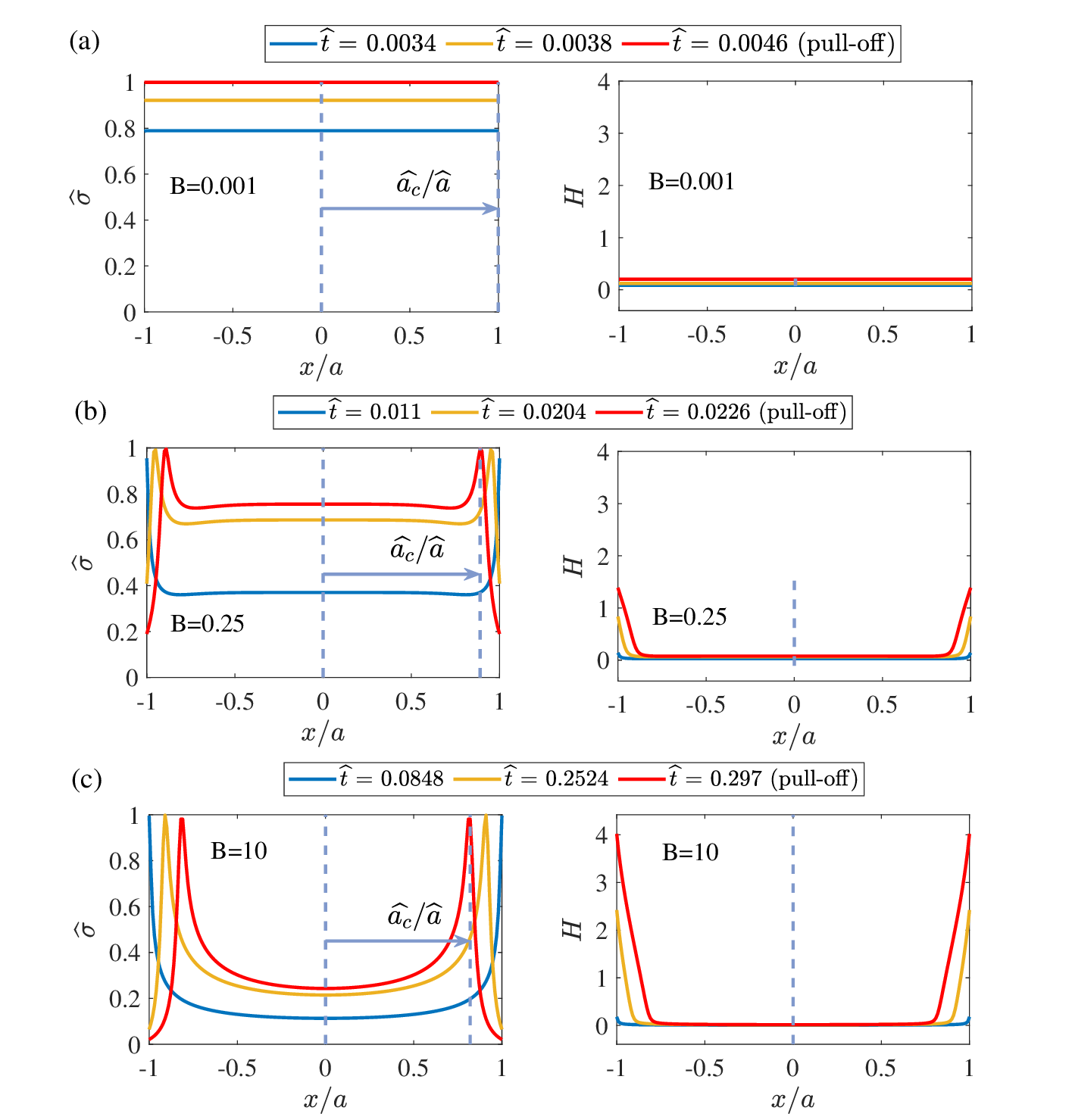}
\caption{{Gap function (the right plots) and the stress distribution (the left plots) on the surface of a layer  with parameters $\Sigma_0 = 0.05$, $k = 0.1$, and $\widehat{a}/\widehat{a}_0 = 64.85$ are presented for various geometries: (a) $B = 0.001$, (b) $B = 0.25$, and (c) $B = 10$, all under an unloading rate of $\widehat{r} = 100$. Each plot displays results for three different moments, with pull-off data highlighted in red.}}
\label{fig:ptilde}
\end{center} 
\end{figure}
A more detailed view on the dependence of the pull-off stress on rate effects is shown in Fig. \ref{fig:stressvsvc} that shows $\widehat{\overline{\sigma}}_{po}$ as a function of the crack speed at pull-off, defined as 
\begin{equation}
V_c=-da_c/dt\;,
\end{equation}
{where $a_c$ represents the crack ligament semi-width, which decreases as the crack propagates at the interface.} The results are closely related to the interaction between adhesion and viscoelastic dissipation in the strip (see also \cite{CPM2021}), indeed it represents one of the major objective of viscoelastic crack propagation theories \cite{Greenwood1981, Schapery, SchaperyII, Schapery2022,Persson2005, persson2017, Persson2021}.  Consequently, we conducted additional analyses to examine this effect. We selected four different cases with a punch radius of $\widehat{a}/\widehat{a}_0= 64.85$ and the corresponding thickness ratios of $\widehat{b}/\widehat{a}_0= [3.24, 6.48, 12.97, 32.42]$, corresponding to the blue, orange, yellow, and purple curves, respectively. We conducted numerical experiments with 20 different unloading rates ranging from $\widehat{r}=0.1$ to $\widehat{r}=100$ to obtain curves representing a wide range of the dimensionless crack velocity $\widehat{V}_c = V_c\tau/h_0$ at pull-off.
\begin{figure}[h]
\centering
\includegraphics[width=4.5in]{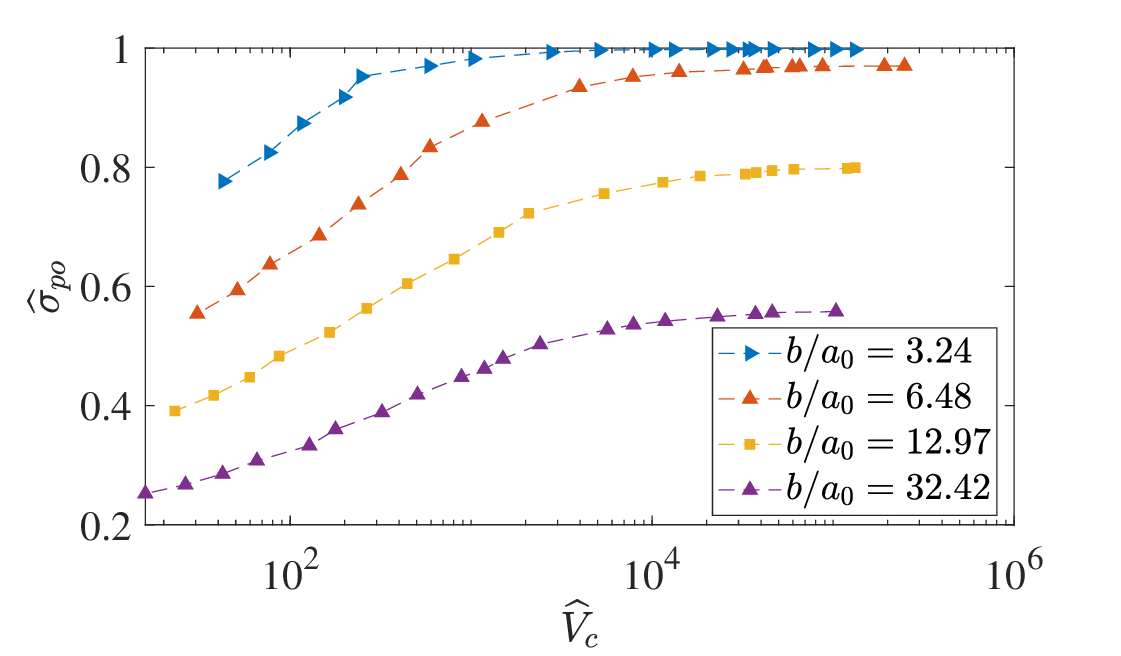}
\caption{Normalized pull-off stress as a function of the normalized crack velocity at pull-off for four different values of $b/a_0$ {for a punch with $\widehat{a}/\widehat{a}_0=64.85$ unloaded from a fully relaxed viscoelastic surface at different unloading rates}.}
\label{fig:stressvsvc}
\end{figure}
{The analysis of Fig. \ref{fig:stressvsvc} illustrates clearly the trend: thin layers and high retraction velocity favour high pull-off stress. Nevertheless, this effect is mitigated when the $b\approx a_0 \approx b_0$ as, in the cohesive region, the detachment tends to happen at a uniform stress.}

\begin{figure}[h]
\centering
\includegraphics[width=4.5in]{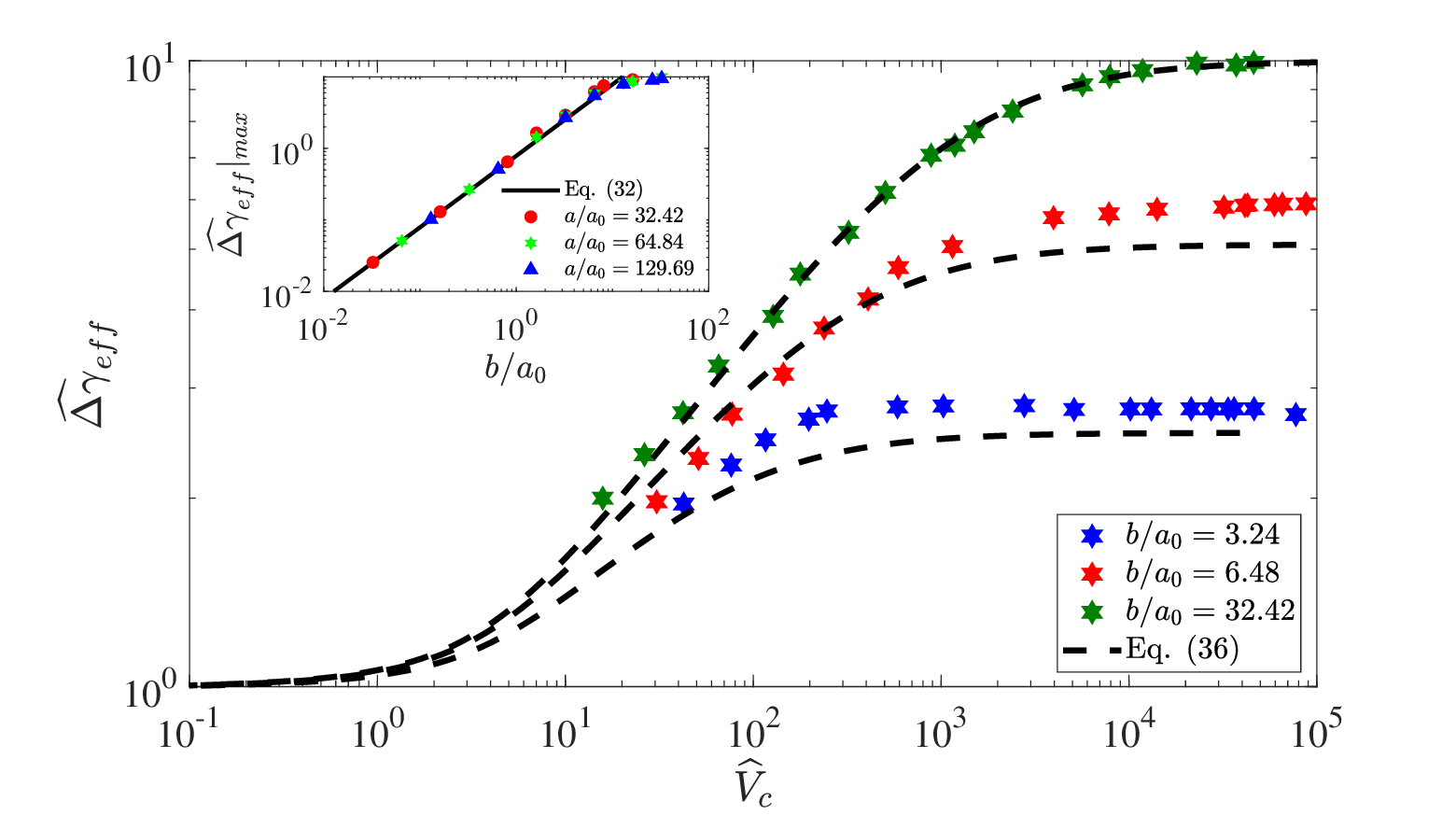}
\caption{Normalized effective surface energy as a function of the crack velocity at pull-off for four different values of $b/a_0$ for a punch with $\widehat{a}/\widehat{a}_0=64.85$ unloaded from a fully relaxed viscoelastic surface at different unloading rates. Dashed lies are obtained with Eq. (\ref{GFinite}). Maximum enhancement for the dimensionless effective surface energy as a function of $b/a_0$. Inset: maximum enhancement of the effective surface energy obtained numerically at very high unloading rates for $a/a_0 = \left[ 32.42, 64.84, 129.69 \right]$, respectively red circles, green stars, blue triangles. The solid black line shows the prediction of Eq. (\ref{dgeff3}).}
\label{fig:Delgammavsvc}
\end{figure}
{Although, the pull-off force reduces increasing the layer thickness the enhancement in terms of effective surface energy remains the same when moving from low to high unloading rates, provided that $b>b_0$. Based on the thin layer elastic solution Eq. (\ref{sl}) we define the effective surface energy as

\begin{equation}\label{dgeffDIM}
{\Delta\gamma}_{eff} =\frac{{\overline{\sigma}}_{po}^2 b }{2 E^*_0}\left(\frac{a}{a_c}\right)^2\;,
\end{equation} 
where we considered that in general the detachment happens at $a_c<a$. Hence, in dimensionless form, 

\begin{equation}\label{dgeff}
\widehat{\Delta\gamma}_{eff} = \frac{\Delta\gamma_{eff}}{\Delta\gamma_0} = \frac{\widehat{\overline{\sigma}}_{po}^2\Sigma_0 \widehat{b} }{2\alpha_{LJ}} \left( \frac{\widehat{a}}{\widehat{a}_c}\right)^2\;.
\end{equation} 

Fig. \ref{fig:Delgammavsvc} shows that the normalized effective surface energy $\widehat{\Delta\gamma}_{eff}$ increases monotonically with respect to the crack velocity $\widehat{V}_c$ at pull-off up to a certain plateau value. In the case of $b/a_0=32.42$, the normalized effective surface energy reaches its theoretical upper limit $\widehat{\Delta\gamma}_{eff}=1/k$ (all our simulations are for k=0.1). Notice that for any thickness of the layer larger than $b_{0\infty}$ one would get the maximum possible enhancement $1/k$. Care should be taken when interpreting the data using Eq. \ref{dgeff} as, if the latter is used for $b>b_1$ this may lead to unrealistic enhancements $\widehat{\Delta\gamma}_{eff}$, which is due to the fact that for $b>b_1$ the halfplane solution should be considered. 

It's important to note that, due to the finite size effect, for cases with $b<b_{0\infty}$, we observe the maximum enhancement of the normalized effective surface energy to be lower than $1/k$. This happens because, for the very thin layer, the cohesive region is approached, namely the DMT-type failure rather than the JKR-type, in the Peng et al. \cite{peng2020decohesion} terminology. {In the latter case, if we assume $\widehat{\overline{\sigma}}_{po}=1$, and $\widehat{a}=\widehat{a}_c$, and with the acquisition of Eq. (\ref{eq:a0}), one can obtain the following relation for the maximum enhancement that can be reached at high retraction rates 
\begin{equation}\label{dgeff3}
\widehat{\Delta\gamma}_{eff}|_{max} = \left(\frac{\pi}{4}\right)\left(\frac{b}{a_0}\right),
\end{equation} 
which turns out to be solely dependent on the ratio $b/a_0$. In order to validate our upper bound enhancement factor (Eq. (\ref{dgeff3})), we considered three distinct values of punch semi-width $a/a_0$, on fully relaxed viscoelastic substrate with varying $b/a_0$ ratios, unloaded at a high unloading rate $\widehat{r}=100$. The inset of Fig. \ref{fig:Delgammavsvc} shows that the maximum enhancement obtained numerically compared very well with Eq. (\ref{dgeff3}).

We incorporated this correction in Greenwood (2004) theory \cite{greenwood2004theory} for crack propagation in viscoelastic semi-infinite media, which, in its original form gives

\begin{equation}
\widehat{\Delta\gamma}_{eff}=\left[  k+\left(  1-k\right)  \frac{\alpha}%
{2}\int_{0}^{1}H\left(  \xi\right)  \exp\left(  -\alpha\left(  1-\xi\right)
\right)  d\xi\right]  ^{-1}\;,\label{GeffG}%
\end{equation}
where
\begin{align}
H\left(  \xi\right)    & =2\xi^{1/2}-\left(  1-\xi\right)  \ln\left(
\frac{1+\xi^{1/2}}{1-\xi^{1/2}}\right)\;,  \\
\alpha & =\frac{\pi}{4\Sigma_{0}}\frac{\widehat{\Delta\gamma}_{eff}\alpha
_{LJ}}{\widehat{V}_{c}}\;.%
\end{align}
Equation (\ref{GeffG}) for very slow propagation gives $\widehat{\Delta\gamma}%
_{eff}=1$, while at high speed provides the maximum enhancement $\left.
\widehat{\Delta\gamma}_{eff}\right\vert _{\max}=1/k$. This picture, on which
all present theories agree, is valid for semi-infinite solids, nevertheless, in agreement with recent results \cite{AffVio2022,VioAffRange,Papangelo2023}, we have found that due to finite size effects the maximum enhancement may be consistently reduced. For the present problem, if $b<b_{0\infty}$, the maximum enhancement will be given by $\left.  \widehat{\Delta\gamma}%
_{eff}\right\vert _{\max}=\left(  \pi/4\right)  \left(  b/a_{0}\right)  $, so
we propose here a generalization of Eq. (\ref{GeffG}) for $b_{0}<b<b_{0\infty}$%

\begin{equation}\label{GFinite}
\widehat{\Delta\gamma}_{eff}\left(  \widehat{V}_{c},\frac{b}{a_{0}}\right)
=\left[  \frac{4}{\pi\left(  b/a_{0}\right)  }+\left(  1-\frac{4}{\pi\left(
b/a_{0}\right)  }\right)  \frac{\alpha}{2}\int_{0}^{1}H\left(  \xi\right)
\exp\left(  -\alpha\left(  1-\xi\right)  \right)  d\xi\right]  ^{-1}\;,%
\end{equation}
where we have explicitly indicated that now the velocity dependent effective surface
energy depends not only on the crack speed, but also on the ratio between the
layer thickness and the fracture length $a_{0}$. Notice that for
$b>b_{0\infty}$ Eq. (\ref{GeffG}) remains valid, while for $b<b_{0}$ the effective energy is
velocity \textit{independent} and equal to $\left.  \widehat{\Delta\gamma
}_{eff}\right\vert _{\max}=\left(  \pi/4\right)  \left(  b/a_{0}\right)$.
Figure \ref{fig:Delgammavsvc} compares the predictions obtained with the finite size Greenwood model (Eq. (\ref{GFinite})) against the numerical results, which are found in fairly good agreement. This result (Eq. (\ref{dgeff3})), obtained with the more general
cohesive-based theory, could also be used to correct the theory of Persson-Brener (2005) \cite{Persson2005} and Persson (2017) \cite{persson2017}, as we shall do in Appendix-II.

One important parameter to examine is the work of separation, which is defined as
\begin{equation}\label{dgeff2}
\widehat{w}_{sep} =\frac{{w}_{sep}}{2 a L \Delta\gamma_0}=\int_{\widehat{\delta}_P}^{\infty} \widehat{\overline{\sigma}} \, d\widehat{\delta}\;.
\end{equation} 
 }which indicates the energy spent during the unloading phase to separate the contact. We calculated this parameter for four different layer thicknesses, specifically $b/a_0 = [3.24, 6.48, 12.97, 32.42]$, which corresponds to the blue, orange, yellow, and purple curves in Fig. \ref{fig:workvsvc}. Similarly to previous research works \cite{AffVio2022,VioAffRange,Papangelo2023} the $\widehat{w}_{sep}$ has a typical bell shape; at low and high velocity there is little energy expenditure to separate the contact as the material behaves essentially as elastic, but for intermediate regimes $\widehat{w}_{sep}$ presents a maximum related to the dissipative phenomena happening in the viscoelastic layer.
\begin{figure}[h]
\centering
\includegraphics[width=4.5in]{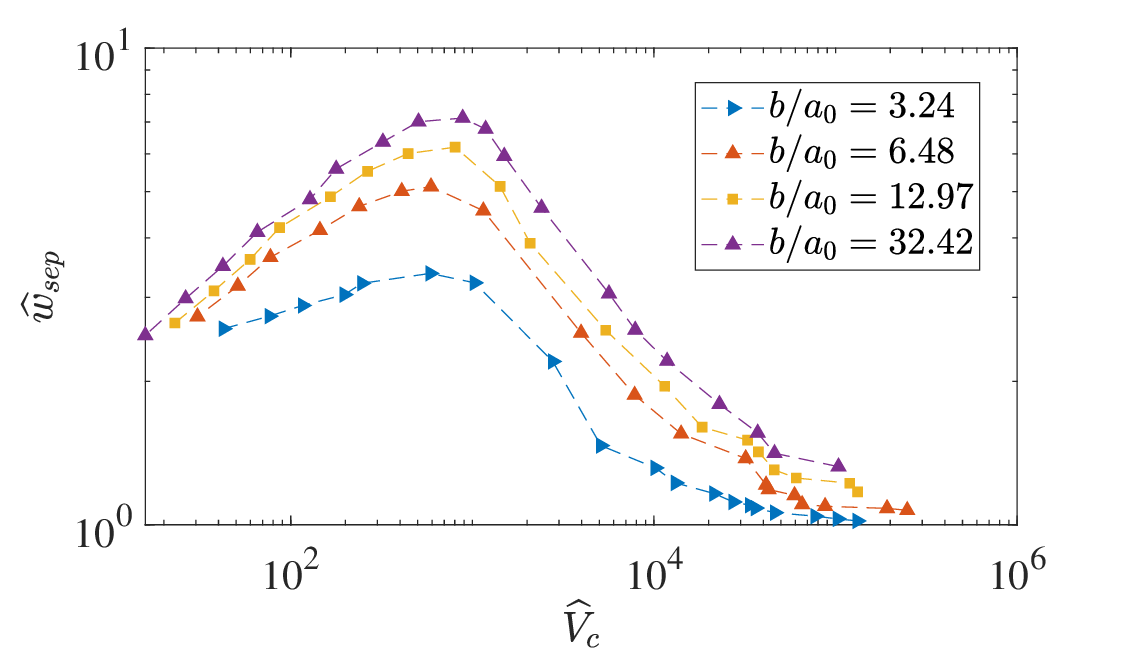}
\caption{Normalized work of separation as a function of the crack velocity at pull-off for four different values of $b/a_0$ {for a punch with $\widehat{a}/\widehat{a}_0=64.85$ unloaded from a fully relaxed viscoelastic surface at different unloading rates}.}
\label{fig:workvsvc}
\end{figure}
}

\section{Conclusions}

The plane problem of the detachment of a large flat punch from an
adhesive viscoelastic layer of finite thickness $b$ has been studied. First,
we have derived an elastic model based on the \textquotedblleft thin
strip\textquotedblright\ assumption by Johnson \cite{Johnson1985}. It was
found that the pull-off stress decays as $\propto 1/\sqrt{b}$. Nevertheless,
this functional dependence is bounded (i) for very thin layer by the
cohesive limit where the pull-off stress equals the theoretical stress of
the material, (ii) for very thick layer by the halfplane limiting solution.
The elastic model provided the bounds for the viscoelastic analysis. We
found that if the layer is thin, particularly at high enough retraction
velocity, the theoretical limit of the material could be reached. This turns
particularly interesting as for soft polymers $E_{\infty }/E_{0}$ may easily
be of the order of $10^{3}\div 10^{4}$ and this amplifies the layer
thickness for which the theoretical strength can be observed. Clearly, this
behavior will be hindered by the fact that during unloading the crack starts
to propagate at the interface hence, at pull-off, the actual crack ligament $%
a_{c}$ is smaller than the punch semi-width.

Theoretical predictions have been compared with boundary element numerical
simulations for a standard linear viscoelastic material and using a
Lennard-Jones force-interaction law. We have shown that the loading
conditions have a negligible effect on the pull-off force, in contrast with
what was shown for a Hertzian geometry. Instead, the pull-off force
consistently increases with the unloading rate up to a certain plateau given
by the cohesive strength of the interface.

Finally, we have shown that when the data are represented in terms of
effective surface energy, at high velocity the theoretical enhancement
given by $E_{\infty }/E_{0}$ is reached only when the layer thickness is
larger than a characteristic lengthscale $b_{0\infty }$. For $%
b_{0}<b<b_{0\infty }$ the maximum adhesion enhancement is limited by finite
size effect and in particular we found $\widehat{\Delta \gamma }%
_{eff}|_{\max }=\left( \pi /4\right) \left( b/a_{0}\right) $. Hence, we have
proposed an extension of Greenwood and Persson crack propagation theories
accounting for finite size effects which we found in good agreement with
numerical results.

\section{Funding}

A.P. and A.M. were supported by the European Union (ERC-2021-STG,
\textquotedblleft Towards Future Interfaces With Tuneable Adhesion By Dynamic
Excitation\textquotedblright\ - SURFACE, Project ID: 101039198, Grant No. CUP:
D95F22000430006). Views and opinions expressed are however those of the
authors only and do not necessarily reflect those of the European Union or the
European Research Council. Neither the European Union nor the granting
authority can be held responsible for them. A.P. was partially supported by
Regione Puglia (Italy), project ENOVIM (Grant No. CUP: D95F21000910002)
granted within the call \textquotedblleft Progetti di ricerca scientifica
innovativi di elevato standard internazionale\textquotedblright\ (art. 22
della legge regionale 30 novembre 2019, n. 52 approvata con A.D. n. 89 of
10-02-2021, BURP n. 25 del 18-02-2021). All the authors acknowledge support by
the Italian Ministry of University and Research (MUR) under the programme
\textquotedblleft Department of Excellence\textquotedblright\ Legge 232/2016
(Grant No. CUP---D93C23000100001).

\section*{Appendices}

{\section*{Appendix-I. Perfectly bonded layer} \label{appendixA}}

For the case of a layer perfectly bonded to the rigid substrate, following
Johnson \cite{Johnson1985}, we only need to correct the previous results for the Poisson
effect with $\zeta=\frac{\left(  1-\nu\right)  ^{2}}{1-2\nu}$ (this requires
$\nu\lesssim0.45$, \cite{Johnson1985}), so that we have%

\begin{equation}
\delta_{po}=\sqrt{\frac{2b\Delta\gamma}{\zeta E^{\ast}}},\qquad\overline{\sigma}
_{po}=\sqrt{\frac{2\zeta E^{\ast}\Delta\gamma}{b}}\;,%
\end{equation}
where $\overline{\sigma}_{po}$ equals the cohesive strength of the material for the layer thickness%

\begin{equation}
b_{0}=\frac{2\zeta E^{\ast}\Delta\gamma}{\sigma_{0}^{2}}=\frac{4}{\pi}\zeta
a_{0}\;.%
\end{equation}

{\section*{Appendix-II. An extension of Persson and Brener viscoelastic crack propagation theory accounting for finite size effects} \label{appendixB}}

Using Persson and Brener theory \cite{Persson2005} for a single relaxation time material gives an implicit equation for the effective adhesion energy \cite{ciavacricrimcmeek}

\begin{equation}
\widehat{\Delta\gamma}_{eff}\left(  \widehat{V}_{c}\right)  =\left[  1-\left(
1-k\right)  \frac{\widehat{\Delta\gamma}_{eff}}{\beta\widehat{V}_{c}}\left(
\sqrt{1+\left(  \frac{\beta\widehat{V}_{c}}{\widehat{\Delta\gamma}_{eff}%
}\right)  ^{2}}-1\right)  \right]  ^{-1}\label{GeffPB}\;,%
\end{equation}
being $\beta=64\Sigma_{0}/\left(  9\sqrt{3}\right)  $. Eq. (\ref{GeffPB}) can be extended to finite size systems by accounting that for a very thin layer the maximum enhancement will be reduced to $\left.  \widehat
{\Delta\gamma}_{eff}\right\vert _{\max}=\left(  \pi/4\right)  \left(
b/a_{0}\right)  $, so we propose here a generalization of Eq. (\ref{GeffPB})
in order to take into account finite size systems, i.e. for $b_{0}%
<b<b_{0\infty}$ %

\begin{equation}\label{eq.persson}
\widehat{\Delta\gamma}_{eff}\left(  \widehat{V}_{c},\frac{b}{a_{0}}\right)
=\left[  1-\left(  1-\frac{4}{\pi\left(  b/a_{0}\right)  }\right)
\frac{\widehat{\Delta\gamma}_{eff}}{\beta\widehat{V}_{c}}\left(
\sqrt{1+\left(  \frac{\beta\widehat{V}_{c}}{\widehat{\Delta\gamma}_{eff}%
}\right)  ^{2}}-1\right)  \right]  ^{-1}\;,%
\end{equation}
where we have explicitly indicated that the normalized effective surface energy depends
not only on the crack speed, but also on the ratio between the layer thickness
and the fracture length $a_{0}$. A comparison between the numerical results and Eq. (\ref{eq.persson}) is shown in Fig. \ref{fig:Persson}. 
\begin{figure}[H]
\centering
\includegraphics[width=4.5in]{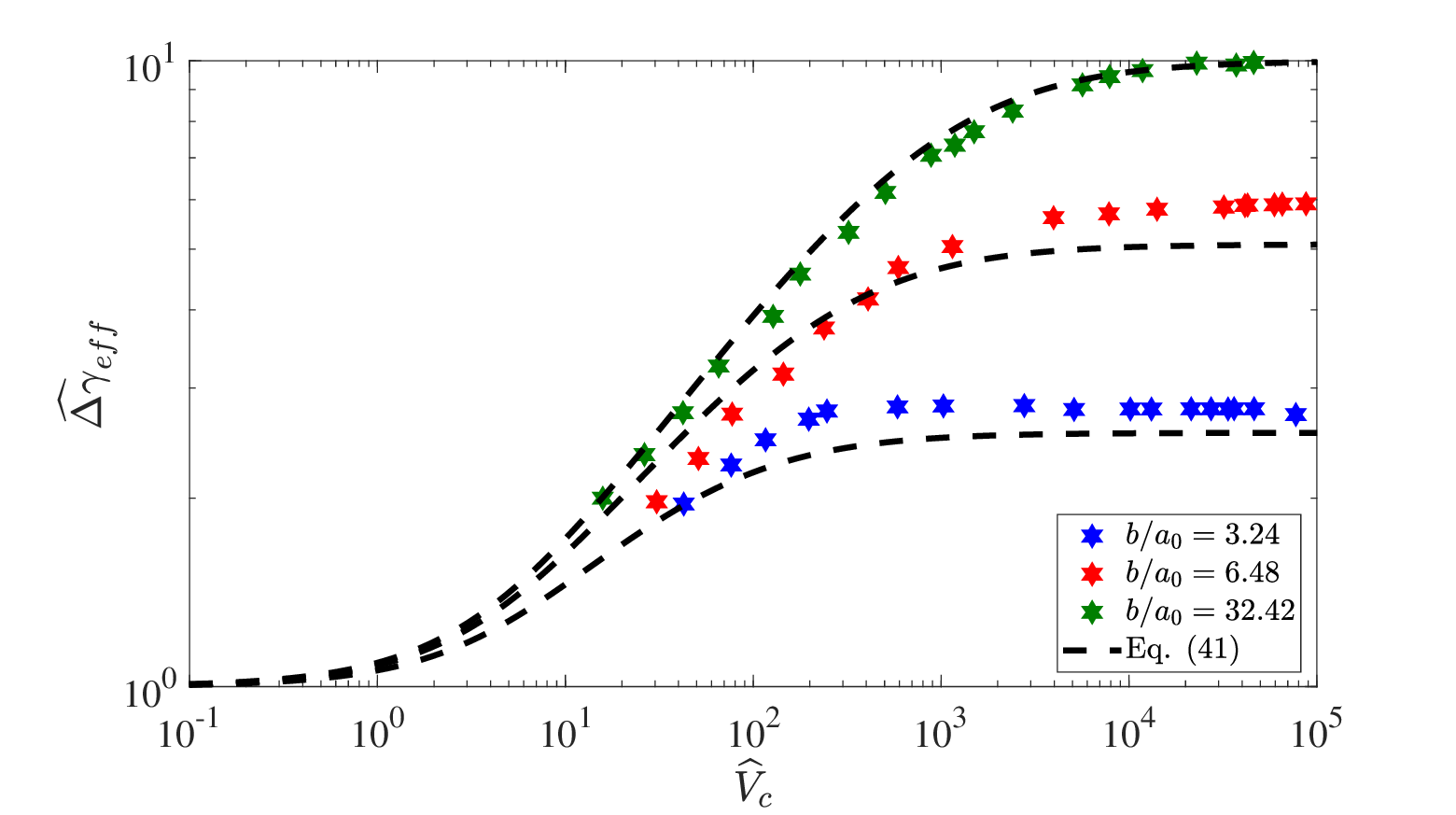}
\caption{Normalized effective surface energy as a function of the crack velocity at pull-off for four different values of $b/a_0$ {for a punch with $\widehat{a}/\widehat{a}_0=64.85$ unloaded from a fully relaxed viscoelastic layer ($k=0.1$)}. Dashed lines are obtained with Eq. (\ref{eq.persson}).}
\label{fig:Persson}
\end{figure}

 \bibliographystyle{elsarticle-num} 
 \bibliography{cas-refs}





\end{document}